\def\half{\frac{1}{2}}
\def\nll{\smallskip\hfil\hfil\linebreak\noindent} 
\def\hi#1#2{$#1$\kern -2pt-#2} 
\def\mb#1{\mbox{\boldmath{$#1$}}}

\def\eq#1{Eq.\,(#1)}
\documentclass[amsmath,amssymb,superscriptaddress]{revtex4}
\usepackage{amsmath}
\usepackage{graphicx}
\usepackage{rotating}

\begin{document}

\vspace*{0.4 in}

\title{Wide effectiveness of a sine basis for quantum-mechanical problems in $d$ dimensions.}
\author{Richard L. Hall}
\email{richard.hall@concordia.ca}
\affiliation{Department of Mathematics and Statistics, Concordia University,
1455 de Maisonneuve Boulevard West, Montr\'eal,
Qu\'ebec, Canada H3G 1M8}
\author{Alexandra Lemus Rodr\'iguez}
\email{a_lemusr@live.concordia.ca}
\affiliation{Department of Mathematics and Statistics, Concordia University,
1455 de Maisonneuve Boulevard West, Montr\'eal,
Qu\'ebec, Canada H3G 1M8}

\begin{abstract}It is shown that the spanning set for $L^2([0, 1])$ provided by the eigenfunctions 
$\{\sqrt{2}\sin(n\pi x)\}_{n= 1}^{\infty}$ of the particle-in-a-box in quantum mechanics provide a very effective variational basis for more general problems. The basis is scaled to $[a,b],$ where $a$ and $b$ are then used as variational parameters. What is perhaps a natural basis for quantum systems confined to a spherical box in $R^d$, turns out to be appropriate also for problems that are softly confined by U-shaped potentials, including those with strong singularities at $r=0.$ Specific examples are discussed in detail, along with some bound $N$-boson systems. 

\end{abstract}
\maketitle
\noindent {\bf keywords:~}Schr\"odinger equation, discrete spectrum, variational analysis, sine basis, confined quantum systems, N-boson problem
\vskip0.2in

\section{Introduction}
We contrast two types of confinement for quantum systems, namely confinement in a finite impenetrable box, and soft confinement by means of a U-shaped potential. The simplest example is provided by  pair of rather different problems in dimension $d=1,$ namely the particle in a box $[-L,\,L]$ and the harmonic oscillator in $R.$  We use the orthonormal basis $\{\phi_i\}_{i = 0}^{\infty}$ of the box problem to approximate states of the oscillator. We shall refer to this basis as a sine basis since the $\{\phi_i\}$ are scaled shifted versions of the eigenfunctions $\{\sqrt{2}\sin(n\pi x)\}_{n= 1}^{\infty}$ for the unit box $x \in [0,\ 1]$. Although the box functions are complete for the Hilbert space $L^2([-L,\,L])$, they cannot represent the oscillator's Hermite functions $\psi_i\in L^2(R)$ exactly.  However, every $\phi_i$ is also a member of the Hilbert space $L^2(R).$  This observation allows us to use the sine functions as variational trial functions for the oscillator. The question remains as to what box size $L$ to use.  This is answered by treating $L$ as a variational parameter and minimizing the upper energy estimates with respect to $L.$  For example, we show in section~2 that by using a sine basis of dimension $N=50,$ and optimizing over $L$, we can estimate the first $10$ eigenvalues $\{1,3,5,\dots,19\}$  of the oscillator $H = -\Delta + x^2$ for $d = 1$ with error less than $10^{-9}$.

 In this paper we demonstrate that for problems which are softly confined, or confined to a box whose size is greater than or equal to $L$, the sine functions indeed provide an effective variational basis.  In section~2 we study the harmonic oscillator and the quartic anharmonic oscillator in $d=1$ dimension. In section~3, we look at spherically symmetric attractive potentials in $R^d,$ such as the  oscillator $V(r)=r^2,$ the atom  $V(r) = -1/r,$ and very singular problems $V(r) = A r^2 + B r^{-4} + Cr^{-6},$ where $\mb{r} \in R^d,$ and $r = \|\mb{r}\|.$ Here we employ a sine basis defined on the radial interval $r\in [a,\,b],$ where $a$ and $b$ are {\it both} variational parameters. In section~4 we study problems that are themselves confined to a  finite box 
 \cite{jaber,ed,hallss,aguil,Michels,varshni,aljaber,fcas,hallc,gusun,agboola}, such as confined oscillators \cite{ed,hallss,aguil} and confined atoms \cite{jaber,hallss,aguil,hallc}.
In section~5 we apply the variational analysis two specific many-boson systems bound by attractive central pair potentials in one spatial dimension.


\section{Problems in $R$}
In order to work in $R$, we first consider the solutions to a particle-in-a-box problem confined to the interval $[0,\,1]$. By applying the transformation $\chi = \left(x-a \right) / \left( b-a \right)$, we shift the box from the interval $[0,\,1]$ to a new interval $[a,\,b]$. This gives us new normalized wave functions
\begin{equation}
\label{eq.1}
\phi_{n}(\chi(x)) = \sqrt{\frac{2}{b-a}} \sin \left( n \pi \left( \frac{x-a}{b-a} \right)\right).
\end{equation}
A special case of this shift is given when the endpoints of the box take the values $a = -L$ and $b = L$, with $L>0$. Then we have the explicit wave functions
\begin{equation}
\label{eq.2}
\phi_{n}(\chi(x)) = \sqrt{\frac{1}{L}} \sin \left( n \pi \left( \frac{x+L}{2L} \right)\right).
\end{equation}

We note that the variational basis $\{\phi_i\}_{n=1}^{\infty}$ is a complete orthonormal set for the space $L^2([a, \,b])\subset L^2(R)$, and a general element $\psi \in L^2([a, \,b])$ can be written as the generalized Fourier series
\begin{displaymath}
\psi = \sum_{i=1}^{\infty}c_{i}\phi_{i},
\end{displaymath}
where $c_{i} = \left(\psi, \phi_{i} \right)=\int_a^b\psi(x)\phi_i(x)dx$, with $i=1,2,\dots$. This justifies the use of this basis for a variational analysis in which the box endpoints $\{a,\,b\}$ are to be used as variational parameters. We shall also use a finite basis $\{\phi_n\}_{n=1}^{N}$ and include one normalization constraint $\sum_{n=1}^N c_n^2 =1$. This constrained minimization of the expectation value $(\psi, H\psi)$ with respect to the coefficients $\{c_n\}$ is equivalent to solving the matrix eigenvalue problem $\mathbf{H}\mathbf{v} = {\mathcal E}\mathbf{v}$, where
\begin{equation}
\label{eq.3}
\mathbf{H} = \left[ \left( \phi_{i}, H \phi_{j} \right) \right].
\end{equation}
By the Rayleigh-Ritz principle \cite{reedsimon} for estimating the discrete spectrum of a self-adjoint Schr\"odinger operator that is bounded below, such as $H$, the eigenvalues ${\mathcal E}_n$ of $\mathbf{H}$ are known to be one-by-one upper bounds ${\mathcal E}_n \geq E_n$ to the eigenvalues of $H$. These bounds can subsequently be further minimized with respect to $a$ and $b$, or with respect to $L$ in case $-a = L = b.$. 


Furthermore, to simplify the variational analysis we use the linearity of the operator $H$, in order to split the matrix $\mathbf{H}$ in two parts as follows
\begin{equation}
\label{eq.4}
\mathbf{H} = \mathbf{K} + \mathbf{P},
\end{equation}
where $\mathbf{K} =  \left[ \left( \phi_{i}, -\Delta \phi_{j} \right) \right]$ represents the kinetic energy component, and $\mathbf{P} =  \left[ \left( \phi_{i}, V(x) \phi_{j} \right) \right]$ represents the potential energy component. The kinetic component will be the same for any potential, in fact, for the basis we have chosen, the matrix will be diagonal, where the non-zero elements depend strictly on the variational parameters and have analytical exact solution, for example, if we use a box with endpoints $\{ -L, L \}$, the diagonal elements of $\mathbf{K}$ are given by $i^{2}\pi^{2} / 4 L^{2}$, for $i=1,2,\ldots$. This reduces the total number of calculations required to estimate the eigenvalues of $\mathbf{H}$.

\subsection{The harmonic oscillator}

It is natural to use the well-known oscillator problem as a test for our variational analysis, since the oscillator is not confined explicitly and moreover its eigenfunctions span $L^2(R).$
We take the scaled one-dimensional harmonic oscillator with Schr\"{o}dinger operator $H = -\Delta + x^2$. The solutions to this problem are 
\begin{eqnarray}
\label{eq.5}
E_{n} & = 2n + 1, \\
\psi_{n}(x) & = c_n H_{n}(x)e^{-x^{2}/2},
\end{eqnarray}
for $n=0,1,2,\ldots$, where $n$ is the corresponding state,  $E_{n}$ the energy of the system, $\phi_{n}$ the wave function, $H_{n}$ the Hermite polynomial of order $n$, and $c_n$ is a normalization constant. 
For this example, we use the basis in Eq.(\ref{eq.2}), with $a = -L$, $b = L,$ to construct the matrix $\mathbf{H}$. Here, $L>0$ is regarded as a variational parameter. 
We then perform a variational analysis using a matrix of dimension $N=50$, minimizing the eigenvalues $\varepsilon_{n}$ over $L \in [5.5,\,8]$. We obtain the results shown in Table~1.
\begin{table}[hbt!]
\caption{Approximation of the energy levels of the harmonic oscillator $H = -\Delta + x^{2}$ in $R$. Here, $n$ represents the energy state, $E$ the exact solution for the energy, $\varepsilon$ is the upper bound for $E$ obtained by the variational analysis, with the eigenvalues $\varepsilon_{n}$ of $\mathbf{H}$ minimized over the box size, and $L$ is the optimal value obtained. The table shows the energies of the first $12$ states, $n=0,1,\ldots,11$}.
\centering
\begin{tabular}{|c|c|c|c|}
  \hline
  n & $E$ & $\varepsilon$ & $L$ \\\hline
  0 & 1 & 1.0000000000 & 6.86 \\ \hline
  1 & 3 & 3.0000000000 & 7.55 \\ \hline
  2 & 5 & 5.0000000000 & 7.09 \\ \hline
  3 & 7 & 7.0000000000 & 7.14 \\ \hline
  4 & 9 & 9.0000000000 & 7.61 \\ \hline
  5 & 11 & 11.0000000000 & 7.49 \\ \hline
  6 & 13 & 13.0000000000 & 6.85 \\ \hline
  7 & 15 & 15.0000000000 & 7.07 \\ \hline
  8 & 17 & 17.0000000000 & 7.27 \\ \hline
  9 & 19 & 19.0000000000 & 7.43 \\ \hline
 10 & 21 & 21.0000000003 & 7.46 \\ \hline
 11 & 23 & 23.0000000017 & 7.49 \\ \hline
\end{tabular}
\label{table.1}
\end{table}
We note that the absolute approximation error is less than $10^{-9}$ for the first $11$ eigenvalues. Furthermore, we obtained an error less than $10^{-5}$ for the first $20$ eigenvalues. If we choose a larger dimension $N$ for the matrix $\mathbf{H}$, the approximations have a smaller errors, and we can calculate $E$ for higher values of $n$ as well, but these results come with a higher computational cost, and can take a long time. 

\begin{figure}[htbp]
\centering
\includegraphics[scale=0.4]{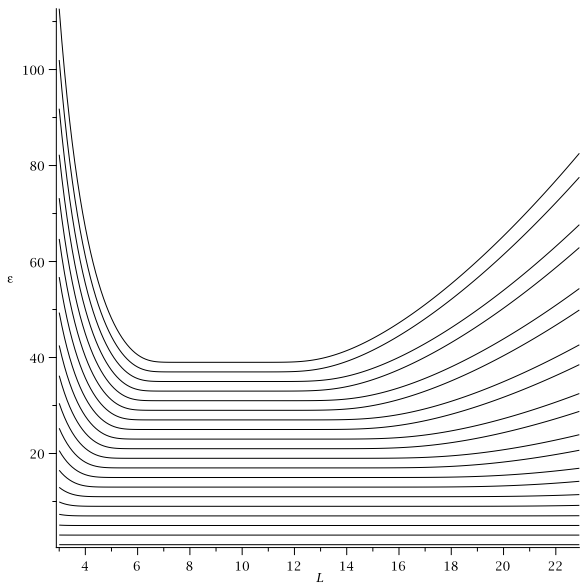}
\caption{Graph of the eigenvalues $\varepsilon$ of the matrix $\mathbf{H}$ as functions of variational parameter $L$ for fixed $N=50$.}
\label{Fig. 1}
\end{figure}

We can consider the energy levels as functions of the parameter $L$ and fixed $N$. Figure (\ref{Fig. 1}) represents the graph of the eigenvalues $\varepsilon_{n}$ of $\mathbf{H}$ versus the variational parameter $L$ with fixed $N=50$ again, for the first $20$ states. We can see that these graphs are $U$ shaped and flat near the minima. If $N$ is large enough, the $U$-shaped graphs become even flatter, this means that if we take any value of $L$ in this flat region, we will end up with good approximations for the energy levels.

We note that for all calculations in this work, we use the computer algebra software {\it Maple.} The advantage of using a program such as this is that it does many of the calculations exactly by using symbolic mathematics; it is only at the end that it resorts to numerical algorithms to solve, for example, the problem of finding the eigenvalues of the Hamiltonian matrix. This minimizes the error obtained in such calculations. 


\subsection{The quartic anharmonic oscillator}

The quartic anharmonic oscillator is another problem in quantum mechanics that has attracted wide interest since Heisenberg studied it in 1925. We consider the special case given by the Hamiltonian $H = -\Delta + x^2 + x^{4}$. Simon \cite{simon} wrote an extensive review of this problem and Banerjee \emph{et al} \cite{anharmonic} calculated the eigenvalues of $H$ using specific scaled basis depending on the harmonic properties of the corresponding eigenfunctions.  Analogously to the previous section, we have performed a variational analysis, in this case using a matrix of dimension $N=20$ and minimizing the eigenvalues $\varepsilon_{n}$ over $L \in [3,\,4]$. The results are exhibited in Table~2.

\begin{table}[hbt!]
\caption{The energy levels for the quartic anharmonic oscillator $H = -\Delta + x^{2} + x^{4}$ in dimension $R$. $n$ represents the energy state, $E$ represents the accurate energy values obtained in \cite{anharmonic}, $\varepsilon$ is the upper bound to $E$ obtained using the present variational analysis in a basis of size $N=20$. The eigenvalues $\varepsilon_{n}$ of $\mathbf{H}$ were minimized  over the box size, and $L$ is the optimal value obtained.  The table shows the first $6$ states, $n=0,1,\ldots,5$.}
\centering
\begin{tabular}{|c |c|c|c|}
  \hline
  n & $E$ & $\varepsilon$ & $L$  \\\hline
  0 & 1.3923516415 & 1.3923516415 & 3.4  \\ \hline
  1 & 4.6488127042 & 4.6488127042 & 3.4  \\ \hline
  2 & 8.6550499577 & 8.6550499586& 3.4  \\ \hline
  3 & 13.1568038980 & 13.1568038994 & 3.7  \\ \hline
  4 & 18.0575574363 & 18.0575574558 & 3.4  \\ \hline
  5 & 23.2974414512 & 23.2974415625 & 3.4  \\ \hline
\end{tabular}
\label{table.2}
\end{table}

We note that with a basis of size only $N=20,$ the approximation error is of the order of $10^{-9}$ for the first five states, and then it grows. This problem is solved by taking a larger $N$ in order to reduce the error.

\section{Problems in $R^d$}


In order to work in higher dimensions where $d > 1$, we need to transform the problem from cartesian coordinates into a more suitable system.  This approach has been studied in depth by Sommerfeld \cite{som}. We let $x= \left(x_{1}, \ldots, x_{d} \right) \in R^{d}$ and transform it into spherical coordinates obtaining $\rho = \left( r, \theta_{1}, \ldots, \theta_{d-1} \right)$ where $ r = \| x \|$. Then the wave function will now be given by
\begin{displaymath}
\Psi(\rho) = \psi(r)Y_{l}(\theta_{1}, \ldots, \theta_{d-1}),
\end{displaymath}
with $\psi(r)$ being the spherically symmetric factor, and $Y_{\ell}$ the hyperspherical harmonic factor, where $\ell = 0, 1, 2,\ldots$.

Given a spherically symmetric potential $V(r)$ in a $d$-dimensional space, using the above tools and following, for example, the work by Hall, \emph{et al.} \cite{ddim}, we get the following radial Schr\"{o}dinger equation
\begin{equation}
\label{eq.6}
-\frac{d^{2} \psi}{dr^{2}} - \frac{d-1}{r} \frac{d\psi}{dr} + \frac{l(l + d - 2)}{r^{2}} \psi + V(r) \psi = E \psi.
\end{equation}
Defining the radial wave function
\begin{displaymath}
R(r) = r^{(d-1)/2}\psi(r),~~R(0)=0
\end{displaymath}
we rewrite equation~(\ref{eq.4}) as
\begin{equation}
\label{eq.7}
-\frac{d^{2}R}{dr^{2}} + UR = ER,
\end{equation}
with effective potential
\begin{equation}
\label{eq.8}
U(r) = V(r) + \frac{(2\ell + d -1)(2\ell + d -3)}{4r^{2}}.
\end{equation}
This analysis allows us to work in higher dimensions whenever we consider spherically-symmetric potentials.

\subsection{The harmonic oscillator}

\begin{figure}[htbp]
\centering
\includegraphics[scale=0.4]{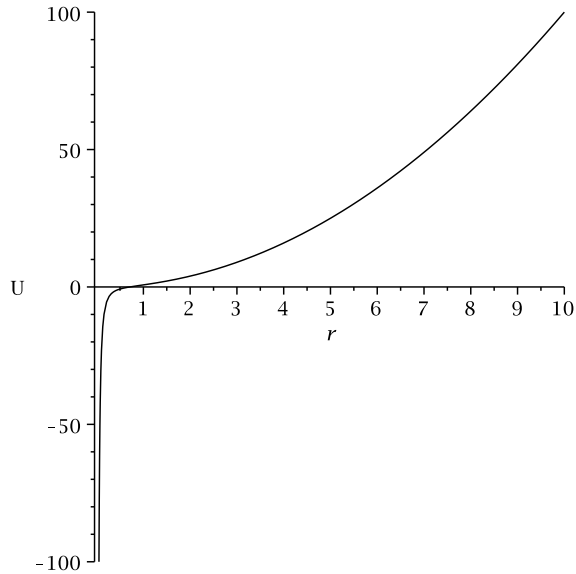}
\caption{Graph of the effective potential $U(r) = r^{2} - 1/4r^{2}$.}
\label{Fig. 2}
\end{figure}

Using the transformation above, we can work with the harmonic oscillator in higher dimensions, $d\geq 2$. A radial Schr\"{o}dinger operator is given now by $H = -\frac{d^{2}}{dr^{2}} + U(r)$, where $U(r)$ is defined as in \eq{\ref{eq.8}} and $V(r) = r^{2}$.
The energy values for this problem are given by
\begin{equation}
\label{eq.9}
E_{n \ell d} = 4n + 2 \ell + d -4,
\end{equation}
where $\ell=0,1,\ldots$ denotes the angular-momentum quantum number for the $d$-dimensional problem. 
The effective potential for this problem has a weak singularity and we have found that the variational basis \eq{\ref{eq.1}} is suitable for such problems, with $a=0$ fixed and $b>0$ as the remaining variational parameter. 
However, we do find some difficulty in dimension $d=2$ when $\ell = 0:$ for this specific case, we obtain the effective potential $U(r) = r^{2} - 1/4r^{2}$. The singular term makes the potential tend to $-\infty$ when $r$ is close to $0$ as shown in Fig. (\ref{Fig. 2}). This is not an inherent feature of the problem  but  indicates a failure of the effective potential representation when  $d=2$ and $\ell = 0:$ the solution to the difficulty is simply to use \eq{7} as the radial differential equation for this particular case. 

We approximate the energy values for the harmonic oscillator in dimensions $d=3,4,5$ and quantum number $l=0,1,2,3$. We use a matrix of dimension $N=40$ and minimize the eigenvalues of $\mathbf{H}$ over $L \in [3,\,12]$. The results are shown in table (\ref{table.3}). 

\begin{table}[hbt!]
\caption{Approximation of the energy levels of the harmonic oscillator in dimensions $d=3,4,5$. $n$ represents the energy state, $E$ the exact solution for the energy given by \eq{\ref{eq.9}}, $\varepsilon$ is the upper bound to $E$ obtained using the present variational analysis. The eigenvalues $\varepsilon_{n}$ of $\mathbf{H}$ are minimized over $L$.} 
\centering
\begin{tabular}{|c|c|c|c|c|c|}
  \hline
  $d$ & $\ell$ & $n$ & $E$ & $\varepsilon$ & $L$ \\ \hline
  3&0 & 1 & 3 & 3.00000000 & 6.00\\ \cline{3-6}
  & &   5 & 19 & 19.00000001 &  7.00\\ \cline{3-6}
  & &   10 & 39 & 39.00000001 & 8.75 \\ \cline{2-6}
  & 1&  1 & 5 & 5.00007348 &  4.50 \\ \cline{3-6}
  &  &  5 & 21 & 21.00167944 &  6.25 \\ \cline{3-6}
  & &   10 & 41 & 41.00907276 &  7.75 \\ \cline{2-6}
   & 2&  1 & 7 & 7.00000000 & 6.00 \\ \cline{3-6}
   & &  5 & 23 & 23.00000001 & 7.75 \\ \cline{3-6}
  & &  10 & 43 & 43.00000001 & 9.25 \\ \cline{2-6}
  & 3 & 1 & 9 & 9.00000001 & 6.00 \\ \cline{3-6}
  & &   5 & 25 & 25.00000076 & 7.25 \\ \cline{3-6}
  & &   10 & 45 & 45.00002070 & 8.50 \\ \hline
  4&0 & 1 & 4 & 4.00073469 & 4.25 \\ \cline{3-6}
  & &   5 & 20 & 20.00745550 &  6.00 \\ \cline{3-6}
  & &   10 & 40 & 40.02454449 & 7.50 \\ \cline{2-6}
  & 1&   1 & 6 & 6.00000262 &  5.00\\ \cline{3-6}
  &  & 5 & 22 & 22.00011370 &  6.50 \\ \cline{3-6}
  & &   10 & 42 & 42.00094014 &  8.00 \\ \cline{2-6}
   & 2&  1 & 8 & 8.00000002 & 6.00 \\ \cline{3-6}
   & &  5 & 24 & 24.00000248 & 7.25 \\ \cline{3-6}
  & &  10 & 44 & 44.00004592 & 8.50 \\ \cline{2-6}
  & 3 &  1 & 10 & 10.00000000 & 6.00 \\ \cline{3-6}
  & &   5 & 26 & 26.00000008 & 7.50 \\ \cline{3-6}
  & &   10 & 46 & 46.00000274 & 8.75 \\ \hline
  5&0 & 1 & 5 & 5.00007348 & 4.50\\ \cline{3-6}
  & &   5 & 21 & 21.00167944 &  6.25\\ \cline{3-6}
  & &   10 & 41 & 41.00907276 & 7.75 \\ \cline{2-6}
  & 1&   1 & 7 & 7.00000000 &  6.00 \\ \cline{3-6}
  &  & 5 & 23 & 23.00000001 &  7.75 \\ \cline{3-6}
  & &   10 & 43 & 43.00000001 &  9.25 \\ \cline{2-6}
   & 2&  1 & 9 & 9.00000001 & 6.00 \\ \cline{3-6}
   & &  5 & 25 & 25.00000076 & 7.25 \\ \cline{3-6}
  & &  10 & 45 & 45.00002070 & 8.50 \\ \cline{2-6}
  & 3 & 1 & 11 & 11.00000001 & 6.00 \\ \cline{3-6}
  & &   5 & 27 & 27.00000000 & 8.00 \\ \cline{3-6}
  & &   10 & 47 & 47.00000001 & 10.00 \\ \hline
\end{tabular}
\label{table.3}
\end{table}

If we increase the dimension of the matrix, we see that the error in the calculations decreases, although the computer time spent increases considerably. Another example is that of approximating the energy values for the harmonic oscillator in dimensions $d=3,4,5$ and quantum number $l=0,1,2,3$ this time for a larger dimension $N$. We used a matrix of dimension $N=500$ and minimized the eigenvalues of $\mathbf{H}$ over $L \in [4,\,12]$. The results are shown in Table~\ref{table.3.1}.

\begin{table}[hbt!]
\caption{Approximation of the energy levels of the harmonic oscillator in dimensions $d=3,4,5$. $n$ represents the energy state, $E$ the exact solution for the energy given by \eq{\ref{eq.9}}, $\varepsilon$ is the upper bound to $E$ obtained using the variational analysis. The eigenvalues $\varepsilon_{n}$ of $\mathbf{H}$ are minimized over $L$. This table shows specific examples of energy values for the quantum numbers $\ell = 0,1,2,3$ and energy states $n=1,5,10$. Note that the approximation error has diminished compared with those of Table~3. In the worst case it is of the of $10^{-5}$, while in others cases the upper bounds are almost exact up to numerical accuracy.}
\centering
\begin{tabular}{|c|c|c|c|c|c|}
  \hline
  $d$ & $\ell$ & $n$ & $E$ & $\varepsilon$ & $L$ \\ \hline
  3&0 & 1 & 3 & 3.00000000 & 5.4\\ \cline{3-6}
  & &   5 & 19 & 19.00000000 &  7.4\\ \cline{3-6}
  & &   10 & 39 & 39.00000001 & 8.7 \\ \cline{2-6}
  & 1&  1 & 5 & 5.000000119 &  5.4 \\ \cline{3-6}
  &  &  5 & 21 & 21.00000234 &  6.9 \\ \cline{3-6}
  & &   10 & 41 & 41.00001129 &  8.4 \\ \cline{2-6}
   & 2&  1 & 7 & 7.000000002 & 6.0 \\ \cline{3-6}
   & &  5 & 23 & 23.00000000 & 9.0 \\ \cline{3-6}
  & &  10 & 43 & 43.00000001 & 9.0 \\ \cline{2-6}
  & 3 & 1 & 9 & 9.000000003 & 6.0 \\ \cline{3-6}
  & &   5 & 25 & 25.00000000 & 8.1 \\ \cline{3-6}
  & &   10 & 45 & 45.00000000 & 10.0 \\ \hline
  4&0 & 1 & 4 & 4.000009387 & 4.7 \\ \cline{3-6}
  & &   5 & 20 & 20.00008831 & 6.5 \\ \cline{3-6}
  & &   10 & 40 & 40.00027064 & 8.1 \\ \cline{2-6}
  & 1&   1 & 6 & 6.000000003 & 7.7\\ \cline{3-6}
  &  & 5 & 22 & 22.00000002 & 7.4 \\ \cline{3-6}
  & &   10 & 42 & 42.00000014 & 8.8 \\ \cline{2-6}
   & 2&  1 & 8 & 8.000000002 & 6.0 \\ \cline{3-6}
   & &  5 & 24 & 24.00000001 & 7.4 \\ \cline{3-6}
  & &  10 & 44 & 44.00000000 & 10.0\\ \cline{2-6}
  & 3 &  1 & 10 & 10.00000000 & 6.6 \\ \cline{3-6}
  & &   5 & 26 & 26.00000000 & 7.6\\ \cline{3-6}
  & &   10 & 46 & 46.00000000 & 10.0 \\ \hline
  5&0 & 1 & 5 & 5.000000119 & 5.4\\ \cline{3-6}
  & &   5 & 21 & 21.00000234 & 6.9\\ \cline{3-6}
  & &   10 & 41 & 41.00001129 & 8.4 \\ \cline{2-6}
  & 1&   1 & 7 & 7.000000002 & 6.0 \\ \cline{3-6}
  &  & 5 & 23 & 43.00000001 & 9.0 \\ \cline{3-6}
  & &   10 & 43 & 43.00000001 &  9.25 \\ \cline{2-6}
   & 2&  1 & 9 & 9.000000003 & 6.0\\ \cline{3-6}
   & &  5 & 25 & 25.00000000 & 8.1\\ \cline{3-6}
  & &  10 & 45 & 45.00000000 & 10.0 \\ \cline{2-6}
  & 3 & 1 & 11 & 1.00000000 & 6.5 \\ \cline{3-6}
  & &   5 & 27 & 27.00000000 & 7.6 \\ \cline{3-6}
  & &   10 & 47 & 46.99999999 & 10.3 \\ \hline
\end{tabular}
\label{table.3.1}
\end{table}

\subsection{The hydrogenic atom}

We consider now a special case of the hydrogenic atom in dimension $d=3$, that is to say a Schr\"{o}dinger operator given by $H = -\frac{d^{2}}{dr^{2}} + U(r)$, with $U(r)$ as in \eq{\ref{eq.8}} and $V(r) =  -\frac{e^{2}}{r} $.
The energy levels for the model hydrogenic atom in this case are given by
\begin{equation}
\label{eq.11}
E_{n\ell}= -\frac{e^{4}}{4(n+\ell)^{2}},
\end{equation}
where $\ell= 0,1,2,\ldots$, and $n=1,2,3,\ldots$.
Since this problem is weakly singular, we use the same basis as in the previous example.  We calculate approximations to the energy values for the case when $e=1$, using a matrix of dimension $N=250$ minimizing the eigenvalues of $\mathbf{H}$ over $L \in [3,\,190]$. The results are shown in Table~\ref{table.4}.

\begin{table}[hbt!]
\caption{Approximation of the energy levels of the hydrogen atom in dimension $d = 3$, for quantum numbers $\ell =0,1,2$. Note that the variational parameter that minimizes the upper bound tends to be very large for all the states. The error is of the order of $10^{-4}$ in the `best' cases.}
\centering
\begin{tabular}{|c|c|c|c|c|c|}
  \hline
  $\ell$ & $n$ & $E$ & $\varepsilon$ & $b$ \\ \hline
  0 & 1 & -0.2500000000 & -0.2499790730, & 17.5\\ \cline{2-5}
  &   2 & -0.06250000000 & -0.06246859682 &  40.5\\ \cline{2-5}
  &   3 & -0.02777777778 & -0.02773301831 & 70.5 \\ \cline{2-5}
  &  4 & -0.01562500000 & -0.01556528040 &  107 \\ \hline
  1 & 1 & -0.06250000000 & -0.06231120892 & 33\\ \cline{2-5}
  &   2 & -0.02777777778 & -0.02747649731 &  60\\ \cline{2-5}
  &   3 & -0.01562500000 & -0.01526320869 & 94 \\ \cline{2-5}
  &  4 & -0.01000000000 &  -0.009656788911 & 143 \\ \hline
  2 & 1 & -0.02777777778 & -0.02777640178 & 75\\ \cline{2-5}
  &   2 & -0.01562500000 & -0.01561644406 &  108\\ \cline{2-5}
  &   3 & -0.01000000000 & -0.009970374676 & 146 \\ \cline{2-5}
  &   4 & -0.006944444444 & -0.006872824074 & 189 \\ \hline
\end{tabular}
\label{table.4}
\end{table}

We see here that the approximation error is larger than $10^{-4}$. There are two problems that arise in this analysis.  First, computations are very slow in this problem due to its singular nature and the number of calculations needed. Second, the hydrogen atom has energy levels that are squeezed together as $n$ grows; meanwhile its wave functions are very spread-out and quite different from those of the particle-in-a-box problem. This confirms what we would expect on general grounds, that the sine basis is not suitable for unconfined atomic problems.

\subsection{Some very singular problems in $R^3$}

Problems involving highly-singular potentials are difficult to solve exactly, but they often provide soft confinement and may be expected to yield to a variational analysis in the sine basis. Test problems are provided by quasi exactly solvable problems.  By this is meant that it is possible to find a part of the energy spectrum exactly provided that some parameters of the potential satisfy certain conditions. Dong \emph{et al}.  \cite{dong} and Hall \emph{et al} \cite{ddim} studied the potential
\begin{displaymath}
V(r) = Ar^{2} + Br^{-4} + Cr^{-6}
\end{displaymath}
in $d=3$. For this work we assume the case where $A=1$, $C>0$ and $\ell = 0$. Then, for this anharmonic singular problem we have the explicit Hamiltonian operator defined by 

\begin{equation}
\label{eq.12}
H = -\frac{d^2}{dr^2} + r^2 +\frac{B}{r^4} +\frac{C}{r^6}. 
\end{equation}
The exact solution for the ground state is given \cite{ddim} by
\begin{equation}
\label{eq.12.1}
E_{0} = 4 + \frac{B}{\sqrt{C}},
\end{equation}
subject to the constraint $(2\sqrt{C}+B)^2 = C(1 + 8\sqrt{C})$. In order to test the sine basis by using a variational analysis for this problem, we considered the exact solutions for the ground state energy in two particular cases: first when $A=B=C=1$ and second, when $A=1, ~B = C = 9$. Since these problems are highly singular, and we are considering radial functions, we need to consider two variational parameters, namely the boundaries of the basis interval, $[a,\,b]$.  Thus, we use the basis given by \eq{\ref{eq.1}} to obtain the matrix $\mathbf{H}$. In this case we need to minimize de eigenvalues with respect to $a>0$ and $b>0$. 
For $A = B = C = 1$ we have the potential
\begin{displaymath}
V(r) = r^{2} + r^{-4} + r^{-6}.
\end{displaymath}
The ground state energy is given by $E_{0} = 5$. We used a matrix of size $N=100$ and found that the best result was the approximation $\varepsilon_{0} = 5.00000003$, with minimizing parameters $a = 0.01$ and $b=5.2$.
For the case where $A=1$ and $B=C=9$ we now have the potential
\begin{displaymath}
V(r) = r^{2} + 9r^{-4} + 9r^{-6}
\end{displaymath}
the ground state energy is given by $E_{0} = 7$. And our approximation is $\varepsilon_{0} = 7.00000110$, where $N=100$, and the variational parameters that give the minimum value are $a=0.01$ and $b=5.1$.  
Even if we have a singular problem, if its potential is $U$-shaped, we can get upper bounds for the energy levels with a small error. For the sine variational basis, the approximations obtained for the upper bound have smaller errors than some of the accurate calculations obtained in the references mentioned above. 

\section{Confined quantum systems}
We can think of this variational approach as if we were confining the system we wish to study in a box, in fact, the same box as the particle-in-a-box problem that generates the basis.  We need only choose the optimal size to find  the best approximations to the energy levels. This opens up the possibility to study confined systems themselves, provided the basis box size $L$ is less than or equal to the size $B$ of the confining box.   The study of these confined quantum systems has been of interest in recent years, for example in the early work of Aguilera-Navarro {\it et al} \cite{aguil}, Michels \cite{Michels}, Ciftci \emph{et al} \cite{hallc}, Al-Jaber \cite{aljaber}, Fernandez and Castro \cite{fcas}.  The sine basis yields upper bounds for the energy eigenvalues for all $L\le B.$  However, we found best results when $L=B.$ This is because the box confinement was dominant for the problems studied. Clearly, with potential confinement and a very large $B$, using an $L$ less than $B$ would be advantageous, as it is for unconfined problems.

\subsection{The confined oscillator}
The confined oscillator was studied by Aguilera-Navarro {\it et al} \cite{aguil} who also used the sine basis,
with basis box size equal to that of the confining box, $L=B.$  We confirm their results, as shown in Table (\ref{table.aguil})
for a box size $B=0.5.$ 
\begin{table}[hbt!]
\caption{Approximation of the energy levels of the harmonic oscillator $H = -\frac{1}{2}\Delta + \frac{1}{2}x^{2}$ in dimension $d=1$. For the state $n$, $E_n$ is a highly-accurate solution for the energy provided by Aguilera-Navarro \emph{et al}. \cite{aguil}, $\varepsilon_n$ is the upper bound for $E_n$ obtained using the present variational analysis with basis size $N=250,$ and $L=0.5$. This table shows the energies of the first $12$ states $n=0,1,\ldots,11$}
\centering
\begin{tabular}{|c|c|c|c|}
  \hline
$n$ & $E$ & $\varepsilon$ \\ \hline
0 & 4.951123323264 & 4.951129323244 \\ \hline 
1 & 19.774534178560 & 19.774534179209 \\ \hline 
2 & 44.452073828864 & 44.452073829725 \\ \hline 
3 & 78.996921150976 & 78.996921150748 \\ \hline 
4 & 123.410710456832 & 123.410710456280 \\ \hline 
5 & 177.693843822080 & 177.693843818558 \\ \hline 
6 & 241.846458758144 & 241.846458765623 \\ \hline 
7 & 315.868612673536 & 315.868612686280 \\ \hline 
8 & 399.760332976128 & 399.760332979135 \\ \hline 
9 & 493.521634054144 & 493.521634068796 \\ \hline 
10 & 597.152524107776 & 597.152524136545 \\ \hline 
11 & 710.653008064512 & 710.653008103290 \\ \hline 
\end{tabular}
\label{table.aguil}
\end{table}

\subsection{The confined sine-squared potential}

Various confined trigonometric potentials have been studied earlier, for example in Refs. \cite{jia,hallp145}. We have found that these problems can be treated very effectively by a variational analysis in the sine basis. We consider one case here, namely the sine-squared potential $V(x)$ given  \cite{hallp145} by
\begin{displaymath}
V(x)= \left\{
\begin{array}{l}
V_0 \sin ^2 (x), ~\hbox{for}~ x \in \left[ -\frac{\pi}{2}, \frac{\pi}{2}\right]\\
\infty, ~\hbox{for}~|x|> \frac{\pi}{2}.\\
\end{array}
\right.
\end{displaymath}
This potential is confined to a box with base of size $\pi$ and height of size $V_0$, as shown in Fig.~3. 
\begin{figure}[htbp]
\centering
\includegraphics[scale=0.4]{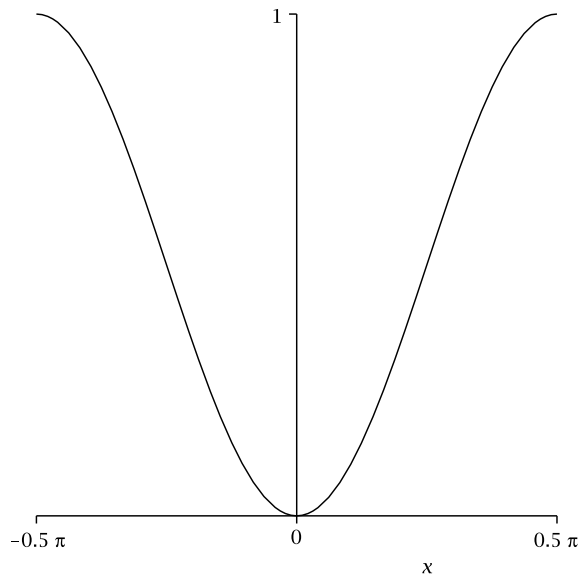}
\caption{Graph of the sine-squared potential for $V_0 = 1$.}
\label{Fig. 3}
\end{figure}

\noindent By using our variational approach, we immediately obtain the energy eigenvalues exhibited in Table \ref{table.sin2} here,  corresponding to those in Table 1 of \cite{hallp145}. For a basis of dimension $N=25$, the result differ by at most $10^{-9}$. We tabulate the relevant results for $V_0 = 0.1, 1, 5$. We have studied both a Hamiltonian Matrix of dimension $N = 10$ and another of dimension $N=25$: the difference in the results between these two variational bases was found to be of order $10^{-13}$ at most for $V_0=0.1,1$ and the order of $10^{-8}$ at most for $V_0 = 5$. 

\begin{table}[hbt!]
\caption{Approximate energy levels for a confined sine squared potential $H = -\Delta + V_0 \sin^2 (x)$ in dimension $d=1$. For the state $n$, $E$ denotes the upper bound for the energies obtained using the present variational analysis with basis size $N=25$ and fixed $L = \frac{\pi}{2}$. The table shows the energies of the first $6$ states $n=0,1,\ldots,5$}
\centering
\centering
\begin{tabular}{|c|c|c|c|}
  \hline
$n$ & $E~(V_0 = 0.1)$ & $E~(V_0 = 1)$  & $E~(V_0 = 5)$ \\ \hline
0 & 1.024922118883 & 1.242428825987  & 2.082985293205 \\ \hline 
1 & 4.049947916808 & 4.494793078632  & 6.370661125009 \\ \hline 
2 & 9.050038818610 & 9.503664867046 & 11.569339156939 \\ \hline 
3 & 16.050020833189 & 16.502081901038 & 18.551201398403 \\ \hline 
4 & 25.050013020839 & 25.501302132228 & 27.532566336109 \\ \hline 
5 & 36.050008928573  & 36.500892873766 & 38.522331587359 \\ \hline 
\end{tabular}
\label{table.sin2}
\end{table}

\subsection{The confined atom}
In the case of the unconfined hydrogen atom we found that we needed ever bigger boxes for each following state because the wave functions are spread-out. However, the present  variational basis is very appropriate for the analysis of confined problems themselves. A hydrogen atom confined to a spherical box has been studied by Varshni \cite{varshni} and by Ciftci, Hall, and Saad \cite{hallc}. In 
\cite{hallc}, the authors found exact solutions for the confined problem given by the Schr\"{o}dinger equation
\begin{equation}
\label{eq.13}
-\frac{d^{2}}{dr^{2}} \psi(r) + \left( \frac{\ell \left( \ell + 1 \right) }{r^{2}} - \frac{A}{r} \right) \psi(r) = E\psi(r),
\end{equation}
with boundary conditions $\psi(0) = \psi(b) = 0$, and $A>0$. These exact solutions are special for the 3-dimensional case.  For different quantum numbers, there are specific radii of confinement for which exact solutions are known. These problems provide ideal tests for the effectiveness of the sine basis.   Details of these exact solutions may be found in \cite{hallc}. We obtain the results shown in Table~\ref{table.confh} for $A=1$ and the radii $b$ required  by the available exact solutions.
\begin{table}[hbt!]
\caption{Approximation of the energy levels of a confined hydrogenic atom in dimension $d = 3$. The angular-momentum quantum number is $\ell$, $n$ is $1$ plus the number of nodes of the radial wave function, $b$ is the radius of the box, $E$ is the exact value of the energy, $E_f$ is the expression in floating point arithmetic, and $\varepsilon$ is the variational approximation obtained from the matrix $\mathbf{H}$ of dimension $N=250$.}
\centering
\begin{tabular}{|c|c|c|c|c|c|c|}
  \hline
  $\ell$ & $n$ & $b$ & $E$ & $E_f$ & $\varepsilon$  \\ \hline
  0 & 1 & 4 & -1/16 & -0.06250000000 & -0.0624999668 \\ \hline
  1 & 1 & 12 & -1/36 & -0.02777777778 &-0.0277777498  \\ \hline
  2 & 1 & 24 & -1/64 & -0.01562500000 & -0.0156250000   \\ \hline
  3 & 1 & 40 & -1/100 & -0.01000000000 & -0.0100000000 \\ \hline
  0 & 1 & $3(3-\sqrt{3})$ & -1/36 & -0.02777777778 & -0.0277777466  \\ \cline{2-6}
    & 2 & $3(3+\sqrt{3})$ &  -1/36 &-0.02777777778 & -0.0277775785 \\ \hline
  1 & 1 & $4(5-\sqrt{5})$ & -1/64 & -0.01562500000 & -0.0156249729  \\ \cline{2-6}
  &   2 & $4(5+\sqrt{5})$ & -1/64 & -0.01562500000 &  -0.0156248833  \\ \hline
  2 & 1 & $5(7-\sqrt{7})$ & -1/100 & -0.01000000000 & -0.0100000000  \\ \cline{2-6}
  &   2 & $5(7+\sqrt{7})$ & -1/100 & -0.01000000000 & -0.00999999997  \\ \hline
  3 & 1 &  36 & -1/144 & -0.006944444444 & -0.006944444438  \\ \cline{2-6}
   &  2 &  72 & -1/144 & -0.006944444444 & -0.006944444431 \\ \hline
\end{tabular}
\label{table.confh}
\end{table}
As opposed to what we found in the case of unconfined atomic models, it is clear that the sine basis is very well suited to the corresponding confined problems.

\section{The $N$-body problem}
We show in this section that the sine basis can also be effective for the  many-body problem.  We consider a system of $N$ identical bosons that are bound by attractive pair potentials $V(x_i - x_j)$ in one spatial dimension.  In units in which $\hbar = 1$ and $m = \half$, the Hamiltonian $H$ for this system,  with the centre-of-mass kinetic energy removed, may be written
\begin{equation}\label{NbodyH}
H = -\left(\sum\limits_{i=1}^{N}\partial_i^2 - \frac{1}{N}\left(\sum\limits_{i= 1}^{N}\partial_i\right)^2\right) + \sum\limits_{1=i<j}^{N}V(x_i-x_j).
\end{equation}
By algebraic rearrangement $H$ may be written in the compact form
\begin{equation}\label{NbodyH2}
H =  \sum\limits_{1=i<j}^{N}\left[-\frac{(\partial_i - \partial_j)^2}{N} + V(x_i-x_j)\right].
\end{equation}
If $\Psi$ is the exact normalized ground state for the system corresponding to the energy $E$, then boson symmetry allows the reduction \cite{Hallp44, Hallp43} to the expectation
 of a one-body operator whose spectral bottom, in turn,  provides an energy lower bound $E_L.$  We have in general,
\begin{equation}\label{EL}
E = (\Psi, H\Psi) = (N-1)\left(\Psi,\left[-2\partial_x^2 + \left(\frac{N}{2}\right)V(x)\right]\Psi\right),
\end{equation}
where $x = (x_1-x_2).$ Thus for the harmonic oscillator $V(x) = cx^2,$ we find immediately that $E_L = c^{\half}(N-1)\sqrt{N},$ which result coincides in this case with the known \cite{Houston, Post53} exact $N$-body solution $E = c^{\half}(N-1)\sqrt{N}.$ 
In order to estimate the ground-state energy from above, we employ a single-product trial function $\Phi$ of the form 
\begin{equation}\label{Nbodyf}
\Phi(x_1,x_2, \dots,x_N) = \prod\limits_i^N \phi(x_i),\quad \phi(x_i) = \sqrt{\frac{2}{a}}\,\cos(\pi x_i/a).
\end{equation}
This wave function vanishes outside a box of volume $a^N$ in $R^N.$  Before we optimize with respect to the box size $a$, we have in general
 $E_U = (\Phi, H\Phi)$, where 
\begin{equation}\label{EU}
E_U = (N-1)\left[\left(\frac{\pi}{a}\right)^2 + \frac{N}{2}\left(\phi(x_1)\phi(x_2)V(x_1-x_2),\phi(x_1)\phi(x_2)\right)\right]
\end{equation}
If we apply Eq.~(\ref{EU}) to the harmonic oscillator $V(x) = cx^2,$  we find
\begin{equation}\label{HOEU}
E_U = (N-1)\min\limits_{a > 0}\left[\left(\frac{\pi}{a}\right)^2 + \frac{Nca^2}{4}\left(\frac{1}{3}-\frac{2}{\pi^2}\right)\right], 
\end{equation}
that is to say,
\begin{equation}
E_U = c^{\half} A(N-1)\sqrt{N},\quad{\rm where}\quad A = (\pi^2/3-2)^{\half}\approx 1.13572.
\end{equation}
Another soluble $N$-boson problem is that of the attractive delta potential $V(x) = -c\delta(x).$ The exact ground-state energy was found by McGuire \cite{McGuire, Mattis} and is given by the formula $E  = -\frac{1}{48}c^2 N(N^2-1).$ Meanwhile the lower and upper bounds we obtain respectively from Eqs.~(\ref{EL}) and (\ref{EU}) are given by
\begin{equation}\label{deltaLU}
E_L = -\frac{1}{32}c^2(N-1)N^2\,  <\, E\, <\, -\frac{9}{64\pi^2}c^2(N-1)N^2 = E_U.
\end{equation}
The lower bound of course agrees with the exact solution for $N = 2.$ For other numbers of particles, the estimates, just as for the corresponding Coulomb one-particle problem, are weaker than those for the tightly bound harmonic oscillator. It is also curious that neither bound manages to reproduce the correct $N$ dependence exhibited by the exact formula of McGuire and Mattis. 

\section{Conclusion}
If we compare the harmonic oscillator $H = -\Delta + r^2$ with the hydrogenic atom $H = -\Delta -1/r$ in three dimensions we see two very different systems from the point of view of stability and the spatial distribution of the respective wave functions.  The oscillator is tightly bound and hardly exists outside a ball of radius $6$, whereas the atom is loosely bound and must be considered out to a radius of $50$ or more. It is therefore not surprising that the more-confined problem, the oscillator, yields to a variational analysis in terms of the sine basis, but the atom does not. The particle in a box is the quintessential confined problem.  It generates a basis that at first sight might appear inappropriate for more general problems. We have shown that it is in fact very effective for problems that are either confined by the nature of the potentials involved or are in any case confined by the given boundary conditions.
For a systems of $N$ identical bosons interacting by attractive pair potentials, the boson permutation symmetry induces behaviour close to that of a scaled two-body problem in which the kinetic-energy term is multiplied by $(N-1)$ and the potential-energy term is multiplied by $N(N-1)/2.$ We show that the ground state of this many-body problem can be  effectively modelled by 
a product of particle-in-a-box wave functions optimized over the box size $L.$

 \section*{Acknowledgement}
Partial financial support of his research under Grant No.~GP3438 from~the Natural
Sciences and Engineering Research Council of Canada is gratefully acknowledged
by one of us (RLH), and one of us (ALR) is grateful for a Doctoral Scholarship from CONACyT (Consejo Nacional de Ciencia y Tecnologia, Mexico).
\nll{\bf References}\smallskip

\end{document}